\documentclass[a4paper]{jpconf}
\usepackage{graphicx}

\begin{document}
%\doublespacing
%\today\\

\title{Development of a new analysis technique
to measure low radial-order p modes in spatially-resolved helioseismic data}

\author{David Salabert$^1$, John W~Leibacher$^{1,2}$ and Thierry Appourchaux$^2$}

\address{$^1$ National Solar Observatory, 950 North Cherry Avenue, Tucson, AZ 85719, USA}
\address{$^2$ Institut d'Astrophysique Spatiale, CNRS-Universit\'e Paris XI UMR 8617, 
              91405 Orsay Cedex, France}

\ead{dsalabert@nso.edu}

\begin{abstract}
In order to take full advantage of the long time series collected 
by the GONG and MDI helioseismic projects,
we present here an adaptation of the rotation-corrected $m$-averaged 
spectrum technique in order to observe low radial-order solar p modes. 
Modeled profiles of the solar rotation demonstrated 
the potential advantage of such a technique \cite{brown85, schou04, app00}. 
Here we develop a new analysis 
procedure which finds the best estimates of the shift of each $m$ of a 
given ($n,\ell$) multiplet, commonly expressed as an expansion in 
a set of orthogonal polynomials, which yield the narrowest mode 
in the $m$-averaged spectrum. We apply the technique to 
the GONG data for modes with $1 \leq \ell \leq 25$ and show that it
allows us to measure lower-frequency modes than with classic 
peak-fitting analysis of the individual-$m$ spectra.
\end{abstract}

\section{Introduction}
In the search for long-lived, low-order p modes, the classic methods 
of peak-fitting the individual-$m$ spectra fail because of the decrease in 
mode amplitudes and increase in solar convective motions (solar noise) 
as the frequency decreases, reducing the signal-to-noise ratio 
(hereafter SNR) of those modes. 
Instead, various pattern-recognition techniques have been developed in 
an effort to reveal the presence of modes in the low-frequency range. 
In the case of spatially-resolved helioseismic data (for example 
observations collected by the space-based SOHO/MDI instrument and 
the ground-based multi-site GONG network), $m$-averaged spectra appeared 
to be a powerful tool. They were employed early in the 
development of helioseismology \cite{brown85}, but have been replaced
by fitting the $m$-spectra individually as the quality of the data improved. 
Years later, in order to fully take advantage of the long-duration 
helioseismic MDI and GONG instruments, 
$m$-averaged spectra corrected by the modeled 
solar rotation were used to detect new low radial-order 
p modes and to put some upper limits on the detectability 
of the g modes \cite{schou04,app00}. 
These authors demonstrated the potential advantage of such 
rotation-corrected $m$-averaged spectra.
%Indeed, a high SNR can result from combining individual low-SNR 
%individual-$m$ spectra, none of which would yield a strong enough 
%peak to fit. 
We present here an adaptation of the $m$-averaged spectrum 
technique in which the $m$-dependent shift parameters are determined.

\section{Observation of long-lived, low signal-to-noise-ratio modes}
The accurate observation of low-frequency modes 
is a difficult task because of two main reasons. First, 
the amplitude of the acoustic modes decreases as the mode inertia 
increases as the frequency decreases, while the ``solar noise'' 
from incoherent, convective motions
increases; thus the SNR of those modes is 
progressively reduced. Second, these low-frequency p modes 
have very long lifetimes, as much as several years. 
%, which means that we need to observe longer to have several excitations
%from the turbulent convection to get a representative 
%measure of the mode properties in the analyzed data.
These two effects together render the observation 
of such low-amplitude and long-lived modes a difficult task to achieve.
However, thanks to several long-duration space- and ground-based instruments, 
more than 11 years of high-quality helioseismic data are available today. 
The observation of low radial-order solar p modes is becoming possible 
if we can somehow improve the SNR of these modes.

The classic peak-fitting analysis consists of fitting 
the $2\ell+1$ individual-$m$ spectra of a given 
multiplet ($n, \ell$), either individually or simultaneously. 
Such fitting methods fail to obtain reliable estimates 
of the mode parameters when the SNR 
is very low. Instead, for a given multiplet ($n, \ell$),
the mean of $2\ell+1$ low-SNR spectra can result in an average spectrum 
with a SNR $\gg$ 1 once the individual-$m$ spectra are corrected for
the rotation- and structure-induced shifts. A model of the solar rotation 
can be used \cite{brown85,schou04,app00} or, as described in this paper, 
we can also determine the shifts as we are searching for 
the low-frequency modes.

\section{Determination of the splitting coefficients and figures-of-merit}
\label{sec:method}
The $m$-averaged spectrum is obtained by finding the best 
estimates of the splitting coefficients, commonly 
called $a$-coefficients, which yield the narrowest peak
in the averaged spectrum. The $a$-coefficients are individually 
estimated through an iterative process. Then, for a given mode ($n,\ell$), 
the frequency shift $\delta\nu_{n \ell m}$ is parameterized by a 
set of coefficients, as:

\begin{equation}
  \delta\nu_{n \ell m} = \sum_{i=1}^{i_{max}}{a_i(n,\ell)\mathcal{P}_{i}^{(\ell)}(m)},
\label{eq:poly}
\end{equation}
\noindent
where $a_i(n,\ell)$ are the splitting coefficients,
and $\mathcal{P}_{i}^{(\ell)}$ corresponds to the
Clebsch-Gordan polynomial expansion \cite{ritz91}. 

%The individual-$m$ spectra are shifted 
%{\it frequency bin by frequency bin} for each of the $a$'s.
%%The guesses of the $a_i$'s, taken as the modeled values during the first 
%%iteration, are updated at each step of the iterative process.
%and used as middle point of the scanned region.  
%Then, in the first step of the first iteration, 
%the coefficient $a_1$ is shifted {\it frequency bin by frequency bin} 
%over a selected interval, while the other $a$'s 
%are kept fixed to their theoretical values. For each shifted bin, 
%the individual-$m$ spectra are corrected from the solar rotation with 
%the corresponding Clebsch-Gordan polynomials, and the mean of 
%these $(2\ell+1)$ rotation-corrected spectra is taken. 

%-------------------------------------------------------------------------
\begin{figure*}[t]
\centering
\includegraphics[width=0.327\textwidth]{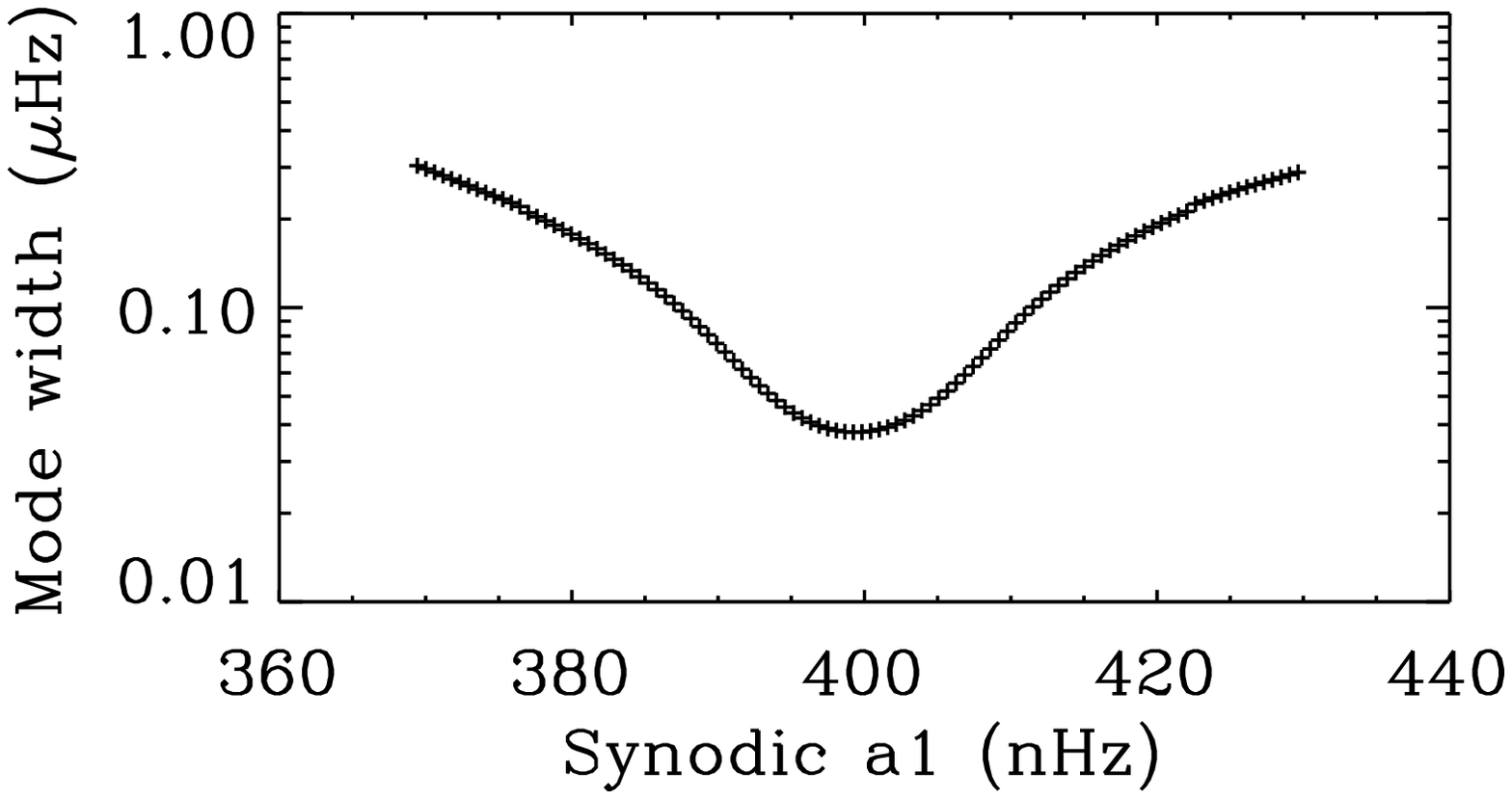}
\includegraphics[width=0.327\textwidth]{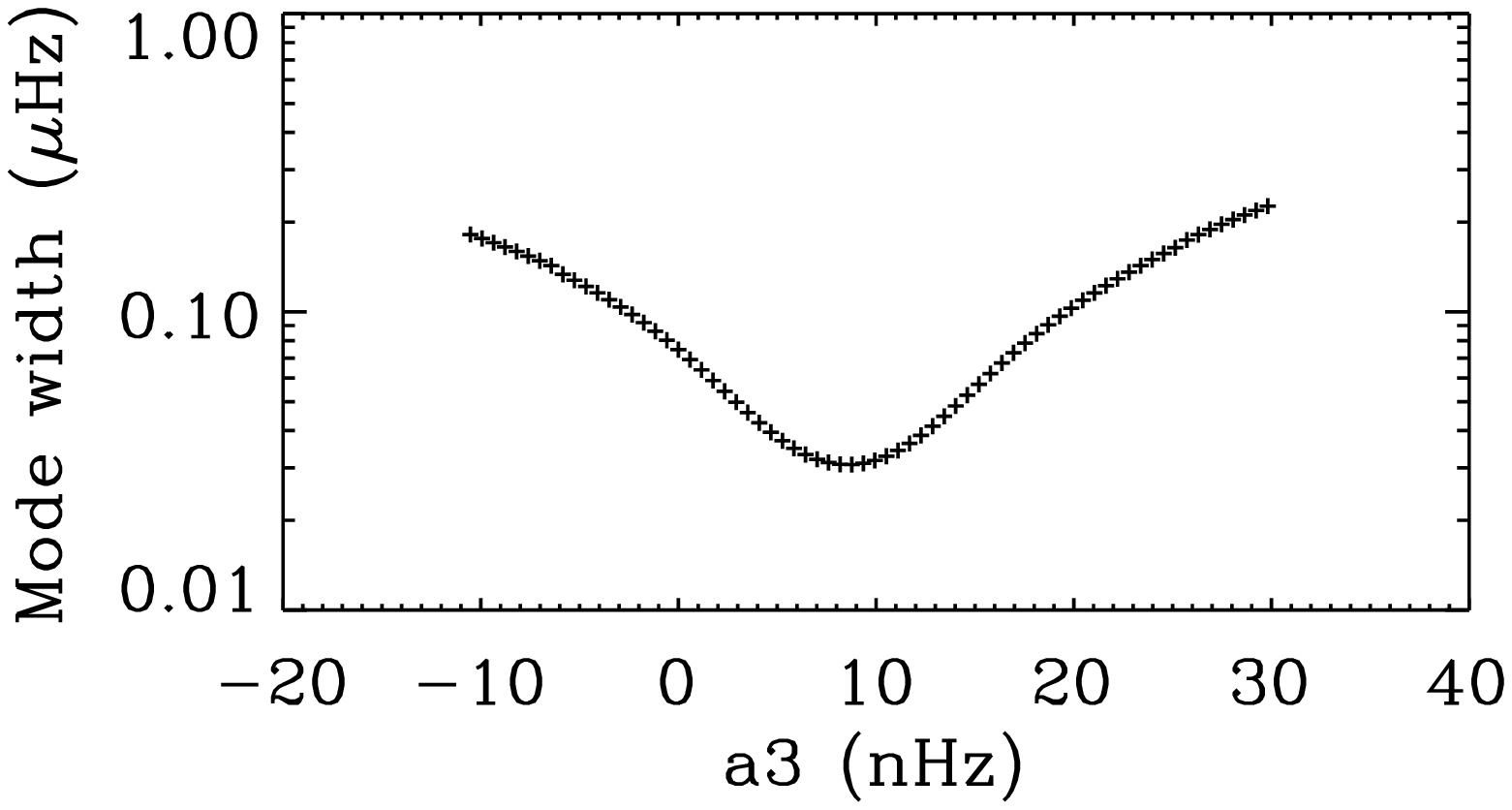}
\includegraphics[width=0.327\textwidth]{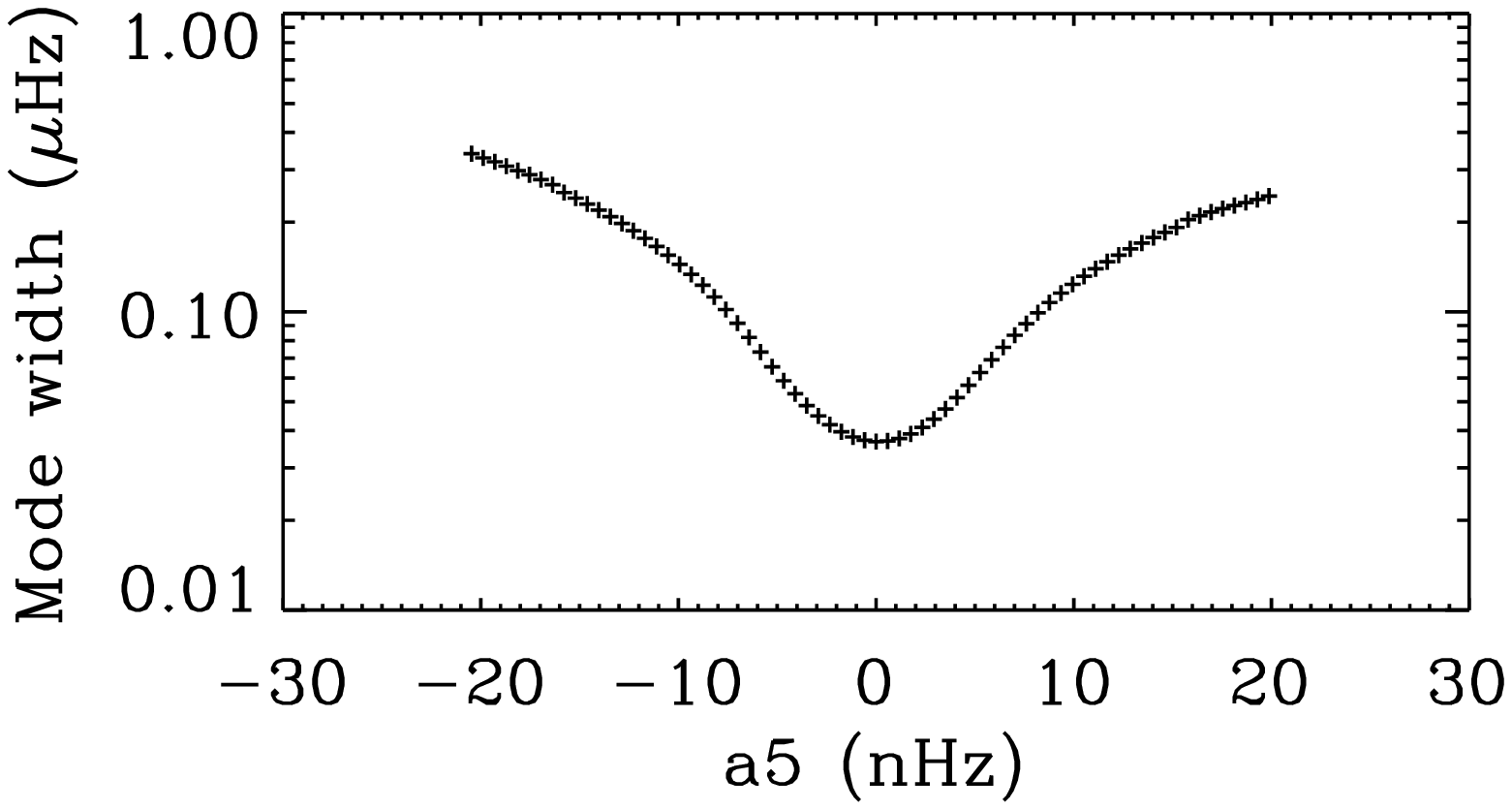}
\caption{Figures-of-merit (FOM) of the oscillation 
multiplet $\ell=5$, $n=6$ showing the fitted peak FWHMs 
of the $m$-averaged spectrum as a function of  
the scanned values of $a_1$, $a_3$, and $a_5$.}
\label{fig:fom}
\end{figure*}
%----------------------------------------------------------------------

For a particular order $i$ of the coefficients $a_i$, a range 
of values is scanned around the current value, 
while the other $a_{i'\neq i}$s are kept fixed. 
For each scanned value of $a_i$, the individual-$m$ spectra are shifted by 
the corresponding Clebsch-Gordan polynomials, and the mean of 
these $2\ell+1$ shifted spectra is taken. 
%The edges of the mean spectrum are apodized by 20\% with a cosine bell function 
%in order to reduce the {\it noise contamination} from the edges of the 
%individual $m$-spectra while correcting for the solar rotation.
The mean of $2\ell+1$ independent 
power densities, which has a $\chi^2$ with more than 2 d.o.f. statistics, 
can be correctly fitted with a Maximum-Likelihood Estimator (MLE) minimization 
code developed for spectra following a $\chi^2$ with 2 d.o.f. statistics
if the SNR of each $m$-spectrum is comparable \cite{app03}.
%if the returned formal uncertainties are normalized by $\sqrt{2\ell+1}$.
%However, this a-posteriori error normalization is correct only
%if the ($2\ell+1$) spectra of given ($n,\ell$) mode have
%the same SNR. The condition of equal SNR among the $m$-spectra of a ($n,\ell$) mode
%being not satisfied in our case, the uncertainties on the mode parameters
%as defined by Appourchaux (2003) have to be taken as first approximations only.
The mean power spectrum is then MLE fitted 
by a Lorenztian profile and its Full-Width-at-Half-Maximum 
(FWHM) determined. When all of the scans are performed, the 
corresponding FWHMs can be seen as a figure-of-merit (FOM). 
As shown on Fig.~\ref{fig:fom}, the FOMs of the $a_i$s clearly
show well-defined minima, the peak profile getting narrower as 
the $a$-coefficients converge to their best estimates.
The uncertainties ($\sigma_{a_i}$) are defined as the formal 
uncertainties of the fits to the minima. 

%The minima of the FOMs become narrower as the mode lifetimes get 
%longer with decreasing frequency. The estimated $a$-coefficients are then
%more precise with sharper FOMs. In order to take into account this
%aspect and being able to estimate the corresponding error bars,
%we decided to perfom a least-squared fit on the FOM's with a Lorentzian 
%profile\footnote{Combined with a power law (Moffat 1969).}. A series of 
%$\chi^2$ tests indicated that a Lorenztian profile describes the FOM's
%better than a Gaussian profile. Thus, the best estimates of the $a_i$s
%correspond to the central frequency of the Lorentzian profile.

%In the following steps, the higher orders of the $a$-coefficients are
%measured in the same manner (Fig.~\ref{fig:fom}), updating at each step 
%the $a$'s estimates. The best set of $a$-coefficients is thus obtained iteratively. 

This iterative procedure is performed until the difference 
between two iterations in each of the computed coefficients $a_i$ 
falls below a given threshold ($10^{-2}$ nHz). A more realistic and 
detailed description of the different effects responsible for lifting the degeneracy of 
the mode frequency into $2\ell+1$ multiplets needs the use of more splitting 
coefficients $a_i$ in Eq.~\ref{eq:poly} than we have shown here, and is currently underway.

Finally, the mode parameters (central frequency, linewidth, $\ldots$) 
are extracted by fitting the $m$-averaged spectrum once the individual-$m$ spectra
are corrected by the best estimates of the splitting coefficients. 
However, the formal uncertainties must be normalized by $\sqrt{2\ell+1}$
\cite{app03}. But this a-posteriori 
error normalization is correct only if the $2\ell+1$ spectra of a 
given ($n,\ell$) mode have the same SNR. The SNRs of the 
$m$-spectra inside a multiplet are not the same in our data, 
and the uncertainties of the mode parameters
have to be taken as a lower limit only.  

%The corresponding 1-sigma errors $\sigma_{a_i}$ 
%are defined as the formal errors returned by the least-squared fit 
%of the FOM functions.

%{\it Since these low-frequency modes have a very low SNR in the 
%individual-$m$ spectra, in order to reduce the number of free 
%parameters, we decided, at the level of development of the analysis
%algorithm so far, to determine only the 3 first odd $a$-coefficients 
%to describe the frequencies of the $m$-components. 
%Even though more $a$-coefficients could have been used, in a 
%first approximation this number
% appeared to be a reasonable compromise between the measurement of the
%solar rotation and the difficulty to determine significant estimates 
%of the rotational splittings for these low-SNR modes. 
%}

%{\it Even if they are significantly non-null, we decided to set to zero the 
%even $a$'s since we actually determined that 
%including the even $a$'s does not significantly improve the mode detection at
%that level of code development.
%Once again, for a more {\it realistic} description of the different
%effects responsible for lifting degeneracy of the mode frequency into
%($2\ell+1$) multiplets, a generalization of the method also includes
%the even $a$-coefficients along with higher orders of the odd $a$'s.  
%}

\begin{figure*}[t]
\begin{center}
\includegraphics[angle=90,width=\textwidth]{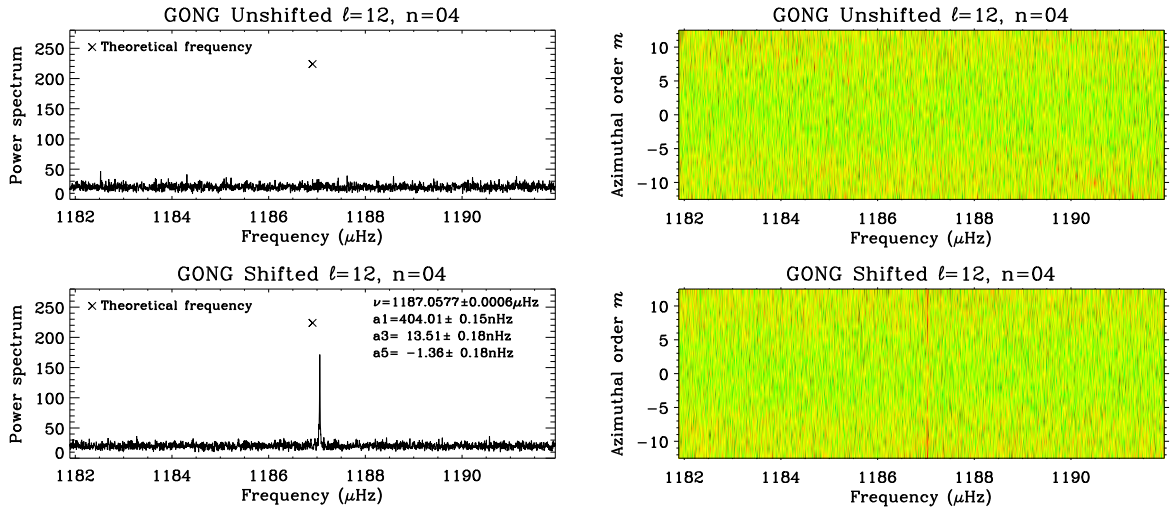}
\end{center}
\caption{Example of $m$-averaged power spectra ({\it left panels})
before ({\it top}) and after correcting for the shift coefficients
({\it bottom}) in the case of the mode $\ell=12$, $n=4$, observed
in 2088 days of GONG data. The estimated central frequencies 
and $a$-coefficients are indicated. The corresponding $m-\nu$ diagrams 
are also shown ({\it right panels}). The crosses indicate the
position of the corresponding theoretical central frequency 
calculated from Christensen-Dalsgaard's model~S.}
\label{fig:spec_l4}
\end{figure*}

\begin{figure*}[h]
\begin{center}
\includegraphics[angle=0,width=\textwidth]{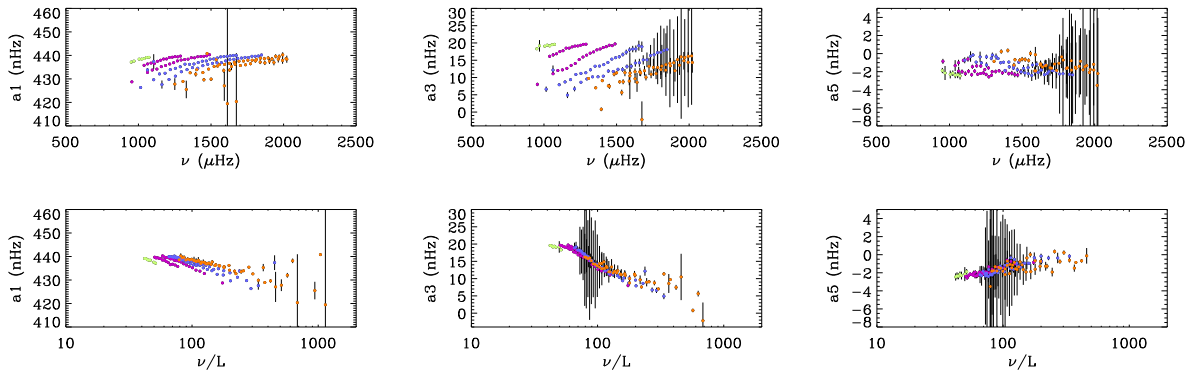}
\end{center}
\caption{Shift coefficients ($a_1$, $a_3$, and $a_5$) of the low
radial-order p modes with $1 \leq \ell \leq 25$ measured in 2088 days of 
GONG data. The $a$-coefficients are represented as a function of 
frequency $\nu$ ({\it upper panels}), and of $\nu/L$, 
with $L=\sqrt{\ell(\ell+1)}$ ({\it lower panels}). The different colors 
correspond to selected ranges of radial order: green dots, modes 
with $n=1,2$; purple, $n=3,4$; blue, $n=5,6$; and orange, $n>6$.}
\label{fig:acoeffs}
\end{figure*}

Figure~\ref{fig:spec_l4} illustrates the advantage of using the 
$m$-averaged spectrum technique in the case of the mode $\ell=12, n=4$
at $\approx$ 1187 $\mu$Hz for 2088 days of GONG data\footnote{Same observational 
timespan as used by \cite{korz05}.}, where 
the $m$-averaged spectra before and after the correction of 
the $m$-spectra for the splitting coefficients are presented. 
%With a proper description of its statistical
%properties, the $m$-averaged spectrum can be fitted 
%as described in Appourchaux (2003), and the corresponding mode parameters 
%(central frequency, linewidth,...) extracted. 
The fitted central frequency (i.e. $a_0$) and 
$a$-coefficients $a_{1,3,5}$ (determined as on Fig.~\ref{fig:fom}) of 
the mode $\ell=12, n=4$ are also given on Fig.~\ref{fig:spec_l4}.
The corresponding $m-\nu$ diagram before correction 
(upper-right panel on Fig.~\ref{fig:spec_l4}) does not show any high 
SNR structure, while after correction (lower-right panel on Fig.~\ref{fig:spec_l4}), 
one can see that the individual-$m$ spectra line up. 
The whole set of estimated shifts $a_1$, $a_3$, and $a_5$ of the low-frequency modes 
with $1\leq\ell\leq25$ and measured in the 2088-day GONG set are 
shown on Fig.~\ref{fig:acoeffs}.

%---------------------------------------------------------------------
%--------------------------------------------------------------------

%---------------------------------------------------------------------
%----------------------------------------------------------------------

\section{Comparison with other measurements}
In the routine analysis, the GONG and MDI projects use 
two separate peak-finding pipelines to extract the mode parameters. 
Those pipelines, developed in the early 1990s and mostly unchanged 
since, provide mode parameters on a routine basis. Time series 
of 108 days are used by the GONG project \cite{anderson90}, 
while the MDI project uses 72-day time series \cite{schou92}.
Recently, a new and independent peak-finding 
method of the individual-$m$ spectra, optimized to take 
advantage of the long, spatially-resolved,
helioseismic time series available today for both GONG and MDI projects,
has been developed \cite{korz05}.

%--------------------------------------------------------------------
\begin{figure}[h]
\includegraphics[width=23pc]{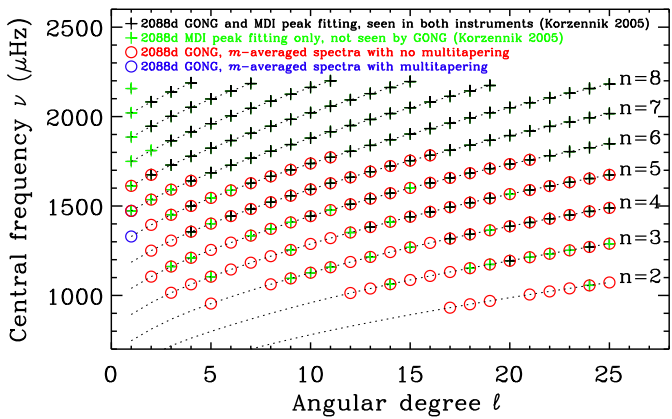}\hspace{2pc}
\begin{minipage}[b]{12pc}\caption{$\ell - \nu$ diagram 
of the low-frequency modes for $\ell=1-25$ measured
in 2088 days of GONG data with the 
$m$-averaged spectrum technique (open circles). The 
crosses correspond to the measurements \cite{korz05} 
with the same 2088 days of GONG and MDI data.}
\label{fig:lnu2088d}
\end{minipage}
\end{figure}
%--------------------------------------------------------------------

This new peak-fitting pipeline has been used 
to extract the low- and medium-degree ($\ell\leq 25$) mode parameters 
on both GONG and MDI observations using one 2088-day long time series, 
as well as using five overlapping segments of 728 days \cite{korz05}. 
In order to compare our results obtained with the 
$m$-averaged spectrum technique, we applied 
the procedure described in Sec.~\ref{sec:method} to the same 
2088 days of GONG observations. 
Only modes which passed the so-called H0 hypothesis 
statistical test \cite{app00} were kept. Figure~\ref{fig:lnu2088d} shows the $\ell-\nu$ 
diagram of the low-frequency modes measured with the two different analysis.
The observed modes with the present method are represented
by the open circles (GONG data only), contrasted with the crosses 
(both GONG and MDI data) \cite{korz05}. 
As seen in Fig.~\ref{fig:lnu2088d}, the $m$-averaged spectrum technique 
allows us 
to measure a significantly larger number of low-frequency modes in GONG data
down to $\approx$ 900 $\mu$Hz, and to fill in the gaps 
in the $\ell-\nu$ diagram obtained by fitting the 
individual-$m$ spectra \cite{korz05}. This $m$-averaged spectrum technique is clearly 
effective in the low-frequency range where classic peak-fitting 
methods are limited by the low signal-to-noise ratio of 
each individual-$m$ component.

\section{Conclusion - Perspectives}
We developed a new method to measure low-frequency 
solar p modes in spatially-resolved helioseismic observations.
The $m$-averaged spectrum technique, in which 
we determine the correction for the splitting coefficients
(i.e., $a$-coefficients), allows us to measure lower 
frequencies in GONG data than with classic peak-fitting methods 
of the individual-$m$ spectra, which are limited by the low
 signal-to-noise ratio of these low-radial order modes. 
We are currently in the process of generalizing the method with a larger
number of $a$-coefficients, both odd and even orders. We are also
going to apply this technique to the MDI data. The lower level 
of background noise in the MDI data should allow us 
to measure even lower low-frequency modes than with GONG observations.

\ack
This work utilizes data obtained by the Global Oscillation Network
Group (GONG) program, managed by the National Solar Observatory, which
is operated by AURA, Inc. under a cooperative agreement with the
National Science Foundation. The data were acquired by instruments
operated by the Big Bear Solar Observatory, High Altitude Observatory,
Learmonth Solar Observatory, Udaipur Solar Observatory, Instituto de
Astrof\'{\i}sica de Canarias, and Cerro Tololo Interamerican
Observatory. This work has been supported by the NASA SEC GIP grant NAG5-11703.

\end{document}